\documentclass[conference]{IEEEtran}

\usepackage{cite}      

\usepackage{graphicx}  

\usepackage{amsmath}   

\usepackage{array}
\begin{document}


\title{Distributed Optimal Quantization and Power Allocation for  Sensor Detection\\ Via Consensus}



 
\author{\IEEEauthorblockN{Edmond Nurellari$^{1}$,  Des McLernon$^1$,  Mounir Ghogho$^{1, 2}$ and Syed Ali Raza Zaidi$^1$}
\IEEEauthorblockA{$^1$University of Leeds, UK \\ 
$^2$International University of Rabbat, Morocco\\ elen@leeds.ac.uk, d.c.mclernon@leeds.ac.uk, m.ghogho@ieee.org, elsarz@leeds.ac.uk }
}

\def\x{{\mathbf x}}
\def\L{{\cal L}}

\maketitle

\begin{abstract}
We address the optimal transmit power allocation problem (from the sensor nodes (SNs) to the fusion center (FC)) for the decentralized detection of an unknown deterministic spatially uncorrelated  signal which is being observed by a distributed wireless sensor network.  We propose a novel fully distributed algorithm, in order to calculate the optimal transmit power allocation for each sensor node (SN) and the optimal number of quantization bits for the test statistic in order to match the channel capacity. The SNs send their  quantized information over orthogonal uncorrelated channels to the FC which linearly combines them and makes a final decision. What makes this scheme attractive is that the SNs share with their neighbours just their individual transmit powers at the current states. As a result, the SN processing complexity is further reduced.
\end{abstract}

%

\section{Introduction}
\label{sec:intro}
Wireless sensor networks (WSNs) are spatially deployed over a field to monitor certain physical or environmental phenomena. Generally, the sensing process is orientated towards estimating various parameters of interest which can be employed to arrive at a certain decision. This decision can then be relayed in a pre-specified manner or can be employed for on-field actuation. We note that the reliable and continued operation of a WSN over many years is often desirable. This is due to the operational environment in which post-deployment access to a sensor node (SN) is at best very limited. Unfortunately, SNs suffer from constrained bandwidth and limited available on-board power. Moreover, due to the locality of the observed process, cooperation amongst SNs is often required to derive an inference. However, such a cooperation comes at the expense of high bandwidth requirements and signalizing overhead. For instance, a WSN formed by $M$ sensor nodes would require transmission of $O(M^2)$ message  exchanges to attain full cooperation. Consequently, designing distributed detection algorithms that efficiently utilize the scarce bandwidth and cope with the impairments in a wireless channel is very important. \\
\indent This work investigates the detection performance of the SN over flat fading wireless transmission links. A centralized solution (taken at the fusion center (FC)) is proposed in \cite{les82} where the deterministic signal ($\boldsymbol s$) to be detected is assumed to be known a-priori. We relax this constraint by deriving a scheme that detects an unknown deterministic signal ($\boldsymbol s$) by employing a linear fusion rule at the FC and adopting the modified deflection coefficient as the detection performance criterion. We also propose a fully distributed algorithm where we allocate the SN transmit power for each individual SN using only local information.   \\
\indent The problem of decentralized detection (and estimation) in a WSN has been extensively tackled in \cite{les82}-\hspace{-0.005cm}\cite{les94}, to name but just a few. Recent publications\cite{Bertrand}-\hspace{-0.005cm}\cite{Lorenzo} propose a distributed algorithm for in-network estimation of algebraic connectivity. Interestingly, \cite{Lorenzo} uses an estimation strategy to adapt the SN transmit power in order to maximize the connectivity of the network, while in this paper we take advantage of the objective function structure and develop a novel distributed algorithm to allocate the SN to FC transmit power. The algorithm is very efficient in terms of convergence and data exchange, also accurate and simple to implement.\\
\indent Section \ref{sec:format} describes the system model and we derive an approach that utilizes the SN to FC channel capacity. An optimum linear combining rule is adopted at the FC with the combining weights optimized in Section \ref{Decentralized optimum weight combining and power allocation}. Section \ref{Decentralized optimum weight combining and power allocation} presents the derivation of the decentralized optimum SN transmit power allocation and our proposed algorithm. Finally, simulation results are given in Section \ref{SIMULATION RESULTS} and conclusions in Section \ref{CONCLUSION}.
%
%
\section{System Model and Quantized decision combining}
\label{sec:format}
Consider the problem of detecting the presence of a deterministic signal $s(n)$ by a sensor network consisting of $M$ SNs. The $i^{th}$ SN collects $N$ samples of the observed signal ($x_i(n)$), and so the detection problem can be formulated as a binary hypothesis test as follows: 
%
%
\begin{eqnarray}
\hspace*{-0.3cm}
\mathcal{ H}_0: x_i\left(n\right) &=&w_i\left(n\right) \\
\mathcal{ H}_1: x_i\left(n\right) &=& s_i\left(n\right)+w_i\left(n\right)
\end{eqnarray}
%
%
where $w_i(n)\sim \mathcal{N}(0,{\sigma_{i}}^{2})$ is AWGN and $s_i(n)$ is the observation of ($s(n)$), both at the $i^{th}$ node. The $i^{th}$ SN then estimates the energy:
%
%
\begin{equation}\label{eq1}
     T_i=\sum \limits_{n=1}^N \left|x_i(n)\right|^{2},  \ i = 1, 2, \ldots, M
\end{equation}
%
%
which for large $N$ can be approximated by a Gaussian distribution \cite{les98} under both hypothesis. So is not difficult to derive (\ref{eq1124})
%
%
\begin{figure*}[h!tb]
\normalsize
\begin{eqnarray}
\mathrm{E}\left\{T_i|\mathcal{ H}_0\right\}=N\sigma_i^{2}, \ \mathrm{Var}\left\{T_i|\mathcal{ H}_0\right\}=2N\sigma_i^{4},\
\mathrm{E}\left\{T_i|\mathcal{ H}_1\right\}\hspace{-0.1cm}=\hspace{-0.1cm}N\sigma_i^{2}\left(1+\xi_i \right),  \ \mathrm{Var}\left\{T_i|\mathcal{ H}_1\right\}\hspace{-0.1cm}=\hspace{-0.1cm}2N\sigma_i^{4}\left(1\hspace{-0.1cm}+\hspace{-0.1cm}2\xi_i \right).
\hspace*{-0.25cm}
\label{eq1124}
\end{eqnarray}
\hrulefill
\vspace*{-12pt}
\end{figure*}
%
%
where $\xi_i =\sum \limits_{n=1}^N  s_i^{2}\left(n\right)/N\sigma_i^{2}$ which can be considered as effective observed SNR. Now linear soft decision combining at the FC has superior performance to the hard decision approach, but it entails additional complexity. In addition soft decision combining puts additional demands on both the limited power resources of the SNs and the effective utilization of the SN to FC channel capacity. So here we propose a scheme, where each individual SN has to quantize its observed test statistic ($T_i$) to $L_i$ bits. The number of quantization bits at the $i^{th}$ SN must satisfy the channel capacity constraint:
%
%
\begin{equation}\label{eq5}
     L_i\leq\frac{1}{2}\log_2\left(1+\frac{p_i h_i^{2}}{\zeta_i}\right) \mathrm{bits}
\end{equation}
%
%
 where $p_i$ denotes the transmit power of sensor $i$, $h_i$ is the flat fading gain between SN $i$ and the FC, and $\zeta_i$ is the  variance of the AWGN at the FC.
The quantized test statistic ($\hat T_i$) at the $i^{th}$ SN can be modeled (with $L_i$ bits) as 
%
%
\begin{equation}\label{eq6}
     \hat T_i=T_i+v_i
\end{equation}
%
%
where $v_i$ is quantization noise (variance, $\sigma_{v_i}^{2}$) independent of $w_i\left(n\right)$ in ((1) and (2)). Assuming quantization noise with a uniform distribution and $T_i\in [0, 2U]$, then
%
%
\begin{equation}\label{eq8}
\sigma_{v_i}^{2}=\frac{U^{2}}{3\times 2^{2L_i}}.
\end{equation}
%
%
Linearly combining $\displaystyle \Big \{  \hat T_i \Big \}_{i=1}^{M}$ at the FC gives
%
%
\begin{equation}\label{eq10}
     T_f=\sum \limits_{i=1}^M \alpha_i \hat T_i
\end{equation}
%
%
where the weights $\displaystyle \big \{  \alpha_i\big \}_{i=1}^{M}$ will be optimized in Section \ref{Decentralized optimum weight combining and power allocation}. Again, for large $M$, $T_f$ will be approximately Gaussian and so we can derive (\ref{eq1122}) and (\ref{eq11}) . We now define $\psi=\mathrm{E}\left\{T_f|\mathcal{ H}_1\right\} -\mathrm{E}\left\{T_f|\mathcal{ H}_0\right\}
=N \sum \limits_{i=1}^M \alpha_i  \left( \sigma_i^{2}\xi_i \right)$ and for a fixed $P_{fa}$ (probability of false alarm) we can write \cite{les96}:
%
%
\begin{figure*}[h!tb]
\begin{eqnarray}
\hspace{-0.3cm}\mathrm{E}\left\{T_f|\mathcal{ H}_0\right\}=\sum \limits_{i=1}^M \alpha_i \left(N\sigma_i^{2}+U\right), \ \mathrm{E}\left\{T_f|\mathcal{ H}_1\right\}=\sum \limits_{i=1}^M \alpha_i \left(N\sigma_i^{2}\left(1+\xi_i \right)+U\right) \hspace{1.3cm}
\label{eq1122}
\end{eqnarray}
\vspace*{-0.4cm}
\begin{eqnarray}
\mathrm{Var}\left\{T_f|\mathcal{ H}_0\right\}=\sum \limits_{i=1}^M \alpha_i ^{2} \left(2N \sigma_i^{4}+\sigma_{v_i}^{2}\right), \
\mathrm{Var}\left\{T_f|\mathcal{ H}_1\right\}=\sum \limits_{i=1}^M \alpha_i ^{2} \left[ 2N  \sigma_i^{4} \left(1+2\xi_i \right) +\sigma_{v_i}^{2} \right]. 
\hspace*{-0.25cm}
\label{eq11}
\end{eqnarray}
\hrulefill
\vspace*{-12pt}
\end{figure*}

%
%
%
%
%
\begin{equation}\label{eq14}
P_d = Q
\left(\frac{Q^{-1}\left(P_{fa}\right) \sqrt {\mathrm{Var} \left\{T_f|\mathcal{ H}_0\right\}} - \psi}{\sqrt {\mathrm{Var}\left\{T_f|\mathcal{ H}_1\right\}}}\right)
\end{equation}
%
%
%
where $P_d$ is the probability of detection.
So using (\ref{eq5}), (\ref{eq8}), (8), (\ref{eq1122}) and (\ref{eq11}) in (\ref{eq14}) we get 
%
%
\begin{multline}\label{eq114}
\hspace{-0.45cm}P_d \hspace{-0.05cm}=\hspace{-0.05cm}Q\hspace{-0.1cm}\left(\hspace{-0.1cm}\frac{Q^{-1}\left(P_{fa}\right) \hspace{-0.1cm}\sqrt { \sum \limits_{i=1}^M \alpha_i ^{2} (2N \sigma_i^{4}+\frac{U^{2}}{3\left(1+\frac{p_ih_i^{2}}{\zeta_i}\right)}) } - \psi}{\sqrt {\sum \limits_{i=1}^M \alpha_i ^{2} [ 2N  \sigma_i^{4} \left(1+2\xi_i \right) +\frac{U^{2}}{3\left(1+\frac{p_ih_i^{2}}{\zeta_i}\right)}]}}\hspace{-0.1cm}\right).
\end{multline}
%
%
The formula in ($\ref{eq114}$) imposes a relationship between the probability of detection, the power allocated to each transmission link (SN to the FC) and the weight ($\alpha_i$ in (\ref{eq10})) for each individual link. 
\section{Decentralized optimum weight combining and power allocation}
\label{Decentralized optimum weight combining and power allocation}
We would now like to find the optimum weighting vector ($\boldsymbol\alpha^o$) and the optimum power allocation vector ($\boldsymbol p^o$) that achieves the best possible $P_d$ (see definitions later), under the constraint of a maximum transmit power budget ($P_t$). However, maximizing  (\ref{eq114}) w.r.t. $\boldsymbol\alpha$ and $\boldsymbol p$ is difficult and no closed form solution can be found. From (\ref{eq114}) it is straightforward to observe that the $P_d$ is a monotonically increasing function of the deflection coefficient. Moreover, $\mathrm{Var}\left\{T_f|\mathcal{ H}_0\right\}<\mathrm{Var}\left\{T_f|\mathcal{ H}_1\right\}$. Employing these two facts, it is intuitive to approximate the optimization problem of (\ref{eq114}) by maximization of the deflection coefficient which is given as:
%
%
\begin{equation}\label{eq17}
     \tilde{d^{2}}\left(\boldsymbol\alpha, \boldsymbol p \right)=\hspace{-0.1cm}\left(\frac{\mathrm{E}\left\{T_f|\mathcal{ H}_1\right\}-\mathrm{E}\left\{T_f|\mathcal{ H}_0\right\}}{\sqrt{\mathrm{Var}\left\{T_f|\mathcal{ H}_1\right\}}}\right)^{2}=\frac{\left(\boldsymbol b^T\boldsymbol  \alpha\right)^2}{\boldsymbol \alpha^T\boldsymbol R_{}\boldsymbol \alpha}\hspace{-0.1cm}
\end{equation}
%
%
where \\ \\
\hspace{-0.1cm}$\boldsymbol b=[N\sigma_1^{2}\xi_1,N\sigma_2^{2}\xi_2,\ldots,N\sigma_M^{2}\xi_M]^T$ \quad \\  \\
$\boldsymbol \alpha=[\alpha_1, \alpha_2, \ldots, \alpha_M]^T$, $\boldsymbol p=[p_1, p_2, \ldots, p_M]^T$ \\ \\
$\boldsymbol R_{}\hspace{-0.1cm}=\hspace{-0.02cm}2N \mathrm{diag}\hspace{-0.02cm}\left(\sigma_1^{4}\left(1\hspace{-0.08cm}+\hspace{-0.08cm}2\xi_1\right)\hspace{-0.05cm}+\hspace{-0.1cm}\frac{\sigma^{2}_{v_1}}{2N},... ,\ \sigma_M^{4}\left(1\hspace{-0.08cm}+\hspace{-0.08cm}2\xi_M\right)\hspace{-0.05cm}+\hspace{-0.1cm}\frac{\sigma^{2}_{v_M}}{2N}\right).\quad$   \\ \\
Note that the dependence of $\tilde{d^{2}}\left(\boldsymbol\alpha, \boldsymbol p \right)$ on the transmit power vector $\boldsymbol p$ enters (\ref{eq17}) through the $\displaystyle \big \{ \sigma^{2}_{v_i}\big \}_{i=1}^{M}$ terms via (\ref{eq5}) and (\ref{eq8}). 
Now, our optimization problem is:
%
%
\begin{equation*}\label{eq20}
\begin{aligned}
\hspace{0.25cm}\left(\boldsymbol \alpha^o, \boldsymbol p^o\right) =\  \underset{\hspace{0.6cm} \boldsymbol \alpha, \boldsymbol p}{\text{arg \ max}}
\hspace{-0.08cm}\left(\tilde{d^{2}}\left(\boldsymbol\alpha, \boldsymbol p \right)\right)  \ \ \ \  \ \ \ \ \ \  \ \ \ \ \ \ \ \ \ \  \\ \quad
\text{\hspace{-2cm}subject to} 
 \sum \limits_{i=1}^M p_i \leq P_t,  \ p_i\geq0,  \ i = 1, 2, \ldots, M. \hspace{0.25cm}
\end{aligned}
\tag{{P1}}
\end{equation*}
%
%
The straightforward solution to (P1) is to obtain it in a centralized manner (i.e., at a FC), where the FC  has full knowledge of the channel gains ($h_i$) which might change over time and need to be updated. The dependence of $\boldsymbol R$ on the flat fading channel coefficients $\left(\displaystyle \big \{ h_i\big \}_{i=1}^{M}\right)$ enters through $\displaystyle \big \{ \sigma^{2}_{v_i}\big \}_{i=1}^{M}$. In this paper we propose a distributed solution, where the SNs are limited to use local information to be able to decide if they should transmit any information to the FC or stay in sleeping mode. 
%
%
\subsection{Optimisation through Decentralized Weight Combining}
\label{sssec:subhead}
Letting $\boldsymbol\beta=\boldsymbol R^{1/2}\boldsymbol \alpha$ in (\ref{eq17}), then we have
%
%
\begin{equation}\label{eq3330}
\tilde{d^{2}}\left(\boldsymbol\beta, \boldsymbol p\right) = \frac{\boldsymbol\beta^T\boldsymbol D\boldsymbol  \beta}{ ||\boldsymbol \beta||^2},\ \boldsymbol D=\boldsymbol {(R_{}}^{-1/2})^T\boldsymbol b\boldsymbol b^{T}\boldsymbol R_{}^{-1/2}
\end {equation}
%
%
and $\boldsymbol \alpha^o=\boldsymbol R_{}^{-1/2}\boldsymbol \beta_{opt}$ in (P1) (assuming $\boldsymbol p$ is constant), where $\boldsymbol \beta_{opt}$ is the eigenvector corresponding to the maximum eigenvalue of $\boldsymbol D$. So we can easily show that:

%
\begin{equation}\label{eq3.111}
\boldsymbol \alpha^o\hspace{-0.01cm}=\hspace{-0.01cm}\left[\hspace{-0.01cm}
       \begin{matrix}
         \frac{N \sigma_1^{2} \xi_1}{2N\sigma_1^{4}\left(1+2\xi_1\right)+\sigma^{2}_{v_1}}, 
                        \cdots, 
\frac{N \sigma_M^{2} \xi_M}{2N\sigma_M^{4}\left(1+2\xi_M\right)+\sigma^{2}_{v_M}}\vspace{0.09cm} 
        \end{matrix}
    \hspace{-0.01cm}\right]
\end{equation}
%
%
Note that (\ref{eq3.111}) establishes a relationship between the optimum weighting vector $\boldsymbol \alpha^o$ and the individual sensor transmit powers through  $\sigma^{2}_{v_i}$ quantity (\ref{eq8}). 
%
%
\subsection{Decentralized Optimum Power Allocation}
\label{sec:pagestyle}
We now propose a novel algorithm aimed at allocating the sensor transmit power to the FC in a fully decentralized fashion. Substitute $\boldsymbol \alpha^o$ from (\ref{eq3.111}) into (P1) to get:
%
%
\begin{equation*}\label{eq202}
\begin{aligned}
& \underset{\boldsymbol p}{\text{\hspace{-0.3cm}maximize}}
& &\hspace{-0.4cm}\left(\hspace{-0.1cm}\sum \limits_{i=1}^M\frac{ N^2 \sigma_i^{4}\xi_i ^{2}}{ {   2N  \sigma_i^{4} \left(1+2\xi_i \right) +\frac{U^{2}}{3\left(1+\frac{p_ih_i^{2}}{\zeta_i}\right)}}}\hspace{-0.1cm}\right) \\ \quad
& \text{\hspace{-0.25cm}subject to} 
& & \sum \limits_{i=1}^M p_i \leq P_t, \ p_i\geq0, \ i = 1, \ldots, M.
\end{aligned}
\tag{{P2}}
\end{equation*}
%
Now (P2) can be solved using the Lagrangian:
%
%
\begin{align*}
   f(\boldsymbol p, \lambda_0, \mu)={ \sum \limits_{i=1}^M \frac{ N^2 \sigma_i^{4}\xi_i ^{2}}{ 2N  \sigma_i^{4} \left(1+2\xi_i \right) +\frac{U^{2}}{3\left(1+\frac{p_ih_i^{2}}{\zeta_i}\right)}} } \hspace{1.5cm}\\ -\lambda_0 \left(\sum \limits_{i=1}^M p_i- P_t\right)+\sum \limits_{i=1}^M \mu_i p_i 
\end{align*}
%
%
and imposing the Karush-Kuhn-Tucker (K.K.T) conditions \cite{Boyd_2}:
%
%
\vspace{-0.5cm}
\begin{center}
\begin{align*}\label{eq21}
 0\in\frac{ N^2 \sigma_i^{4}\xi_i ^{2}}{ \left(\hspace{-0.14cm}2N  \sigma_i^{4} \left(1+2\xi_i \right) +\frac{U^{2}}{3\left(1+\frac{p_ih_i^{2}}{\zeta_i}\right)}\hspace{-0.1cm}\right)^{\hspace{-0.22cm}2}  }\hspace{-0.07cm}\times\hspace{-0.07cm} \frac{U^{2}\times \frac{h_i^{2}}{\zeta_i}}{3\left(1+\frac{p_ih_i^{2}}{\zeta_i}\right)^{\hspace{-0.1cm}2}} \hspace{-0.05cm}-\hspace{-0.05cm}\lambda_0 \hspace{-0.05cm}+\hspace{-0.05cm}\mu_i 
\tag{22}
\end{align*}
\end{center}
%
%

\vspace{-0.7cm}
\begin{align*}\label{eq22}
\lambda_0 \left(\sum \limits_{i=1}^M p_i- P_t\right)=0 \hspace{-0.8cm}\\
\sum \limits_{i=1}^M p_i- P_t \leq0 \hspace{-0.8cm}\\
\lambda_0\geq 0 , \ \mu_i p_i=0,  \ i = 1, 2, \ldots, M \hspace{-0.6cm} \\
\mu_i\geq0, \  p_i\geq0,  \ i = 1, 2, \ldots, M. \hspace{-0.6cm}
\tag{23}
\end{align*}
%
%
We can let the Lagrangian $f(\boldsymbol p, \lambda_0, \mu)$ = $\sum \limits_{i=1}^M f_i(p_i, \lambda_0)$ =
%
%
\begin{align*}
{\sum \limits_{i=1}^M ( {\frac{ N^2 \sigma_i^{4}\xi_i ^{2}}{ {   2N  \sigma_i^{4} \left(1+2\xi_i \right) +\frac{U^{2}}{3\left(1+\frac{p_ih_i^{2}}{\zeta_i}\right)}}}\hspace{-0.1cm}}  -\lambda_0 p_i+ \frac{\lambda_0}{M} P_t)}.
\end{align*}
%
%
Now, (P2) is converted into $M$ separable  problems that can be solved in parallel using the dual ascent algorithm:
%
%
\begin{equation*}\label{eq30}
\vspace{-0.1cm}
\begin{aligned}
\hspace{-3cm}p_i\left[k+1\right]=\underset{\hspace{0.6cm} p_i}{\text{arg \ min}}\hspace{0.1cm}f_i(p_i, \lambda_0\left[k\right]) \\ 
\end{aligned}
\tag{{a1}}
\vspace{-0.1cm}
\end{equation*}
%
%
%
%
\begin{equation*}\label{eq30}
\begin{aligned}
 &\lambda_0\left[k+1\right]=\lambda_0\left[k\right]+\epsilon\left[k\right]\left(\sum \limits_{i=1}^M p_i\left[k+1\right]-P_t\right).
\end{aligned}
\tag{{a2}}
\end{equation*}
%
%
For this formulation we can see that the only step that requires an exchange of values among the sensors is the (a2) step which requires the computation of $\sum \limits_{i=1}^M p_i\left[k+1\right]$ at each sensor node. Because of the communication topology for the $M$ SNs (i.e., not fully connected), we will use the average consensus algorithm \cite{saber} to ensure the availability of this term at each SN. In this paper, we assume ideal exchange of information between sensors that are connected. Solving the K.K.T conditions in $ (\ref{eq21})$ and $(\ref{eq22})$ gives a solution for the optimum $p_i$:
%
%
\begin{align*}\label{eq24}
p^o_{i}\hspace{-0.1cm}=\hspace{-0.1cm}\Biggl[\hspace{-0.1cm}\frac{1}{\sqrt{\lambda_0}}\Biggl(\frac{\xi_i U\sqrt{3}}{6\sigma_i^{2}\left(1\hspace{-0.07cm}+\hspace{-0.07cm}2\xi_i\right)\sqrt{\frac{h_i^{2}}{\zeta_i}}}\Biggl)- \frac{U^2}{6N\sigma_i^{4}\left(1\hspace{-0.07cm}+\hspace{-0.07cm}2\xi_i\right)\frac{h_i^{2}}{\zeta_i}} -\frac{\zeta_i}{h_i^{2}}\hspace{-0.05cm}\Biggr]^{+} 
\tag{24}
\end{align*}
%
%
where
\begin{equation*}\label{eq2254}
[x]^+=\left\{
\begin{aligned}
 x, \ \ \ \  \ \mathrm{ if} \ x\geq0 \ \quad \\
  0, \ \ \ \  \ \mathrm{ if} \ x<0 .\quad 
\end{aligned}
\right. 
\end{equation*}
As mentioned before, the centralized solution at the FC requires full knowledge of the channel gains ($h_i$) which might be time-varying and need to be  always updated. It also requires the variance of AWGN ($\zeta_i$) and each of the local SNRs ($\xi_i$). Moreover, the FC has to broadcast back to each individual SN the allocated SN transmit power which might be decoded with error due to fading. Furthermore, when the FC is battery operated, the centralized solution (at the FC) becomes inefficient and not scalable as the number of SNs increases. On the other hand, the proposed distributed algorithm ($\boldsymbol {Algorithm1}$) is fully scalable in terms of data exchange and SN processing complexity. As is shown in the simulation result it is also very accurate.
We now define $\epsilon[k]$ to be the positive user defined step size and $\overline{p_i\left[k+1\right]}=\frac{1}{M}\sum \limits_{i=1}^M p_i\left[k+1\right]$.
%
%
\begin{table}[htp!] 
\small
\centering 
\begin{tabular}{l r} 
\hline\hline 
$\boldsymbol{Algorithm1:}$ Optimizing the sensor transmit power 
 \\ [0.6ex] 
\hline 
STEP 1:  Set $k=0$, $\kappa$ equal to a small positive value  \\  \hspace{1.2cm} and initialize $\lambda_0\left[0\right]$, $\forall i$;\\  \\ 
STEP 2:  Compute  $p_i\left[1\right]$,  $\forall i $ using (a1);\\  \\
STEP 3:  Run consensus over $p_i\left[1\right]$ to get $\overline{p_i\left[1\right]};$\\ \\ 
STEP 4:  Compute $\lambda_0\left[1\right]$ using (a2);\\ \\
STEP 5:  Set $k=1$;\\ \\
STEP 6:  Repeat until convergence  \\ \\ $p_{i}[k\hspace{-0.07cm}+\hspace{-0.07cm}1]\hspace{-0.07cm}=\hspace{-0.1cm}\Biggl[\hspace{-0.07cm}\frac{1}{\sqrt{\lambda_0[k]}}\Biggl(\hspace{-0.05cm}\frac{\xi_i U\sqrt{3}}{6\sigma_i^{2}\left(1+2\xi_i\right)\sqrt{\frac{h_i^{2}}{\zeta_i}}}\hspace{-0.05cm}\Biggl) - \frac{U^2}{6N\sigma_i^{4}\left(1+2\xi_i\right)\frac{h_i^{2}}{\zeta_i}} \hspace{-0.05cm}-\hspace{-0.05cm}\frac{\zeta_i}{h_i^{2}}\hspace{-0.07cm}\Biggr]^{+}$ \hspace{-2.4cm}\\ \\ Run consensus over $p_i\left[k+1\right] $ until convergence \\ \\ \hspace{0.6cm} $\lambda_0\left[k+1\right]=\lambda_0\left[k\right]+\epsilon\left[k\right]\left(M \overline{p_i\left[k+1\right]}-P_t\right)$ \\ \\ Set $k=k+1$, if convergence criterion is satisfied stop, \\  otherwise go to step 6. \\ [0.5ex]  
\hline
\end{tabular} 
\label{tab:hresult} 
\end{table} 
%
%
The convergence criteria that we use in here is the relative absolute difference: $\frac{||\boldsymbol p[k+1]-\boldsymbol p[k]||}{||\boldsymbol p[k]||}\leq \kappa$, where $\kappa$ is a positive small constant and $\boldsymbol p[k]$ is the vector of the SN transmit power at the $k^{th}$ iteration.\\
 \indent It is also possible to exchange among the SNs the $h_i$, $\zeta_i$, $\sigma_i^2$, and $\xi_i$ quantities $\forall i$ where each SN will store them  in the corresponding vectors $\boldsymbol h$, $\boldsymbol{\zeta}$, $\boldsymbol {\sigma}^2$, and $\boldsymbol{\xi}$ together with their corresponding SN index. When all the quantities will be available at each SN they can be used to allocate the SN transmit power through (\ref{eq24}) and $\lambda_0$ can be calculated through the constraint in (P2). 
%
%
\section{Simulation results}
\label{SIMULATION RESULTS}
In this section, the proposed algorithm is evaluated numerically and compared to its centralized counterpart. Also, we choose $\lambda_0[0]=10^{-8}$, $\forall i$, $\kappa=10^{-7}$ and $\epsilon\left[k\right]= \lambda_0[k]/k$. We let all the  $\sigma_i^{2}$ terms at each SN be different, such that $\xi_a=10\log_{10}\left(\frac{1}{M}\sum \limits_{i=1}^M \xi_i \right)=$ -4 dB, unless otherwise stated. In addition we let $\zeta_i=0.1$ $\forall i$. We compare the results with the matched filter detector\footnote{The test statistic is taken as: $T_i=\sum \limits_{n=1}^N x_i(n)s_i(n),  \ \forall i = 1, 2, \ldots, M$. The global test statistic ($T_f$) at the FC has the same structure as (8) with $\alpha_i=\frac{\sum \limits_{n=1}^N s^2_i(n)}{\sigma^2_i\sum \limits_{n=1}^N s^2_i(n)+\sigma^{2}_{v_i}},\ \forall i = 1, 2, \ldots, M.$ The optimum weights have been derived through the Likelihood Ratio Test (LRT).} (MFD) and use this as a benchmark. We will also refer to ``equal linear combining" in (8) (i.e., $\alpha_i=\frac{1}{\sqrt{M}}, \forall i$) and ``equal power allocation" in (5) (i.e., $p_i=\frac{P_t}{{M}}, \forall i$).
%
%
\begin{figure}[!htp]
 \centerline{\includegraphics[width=80mm ,height=70mm]{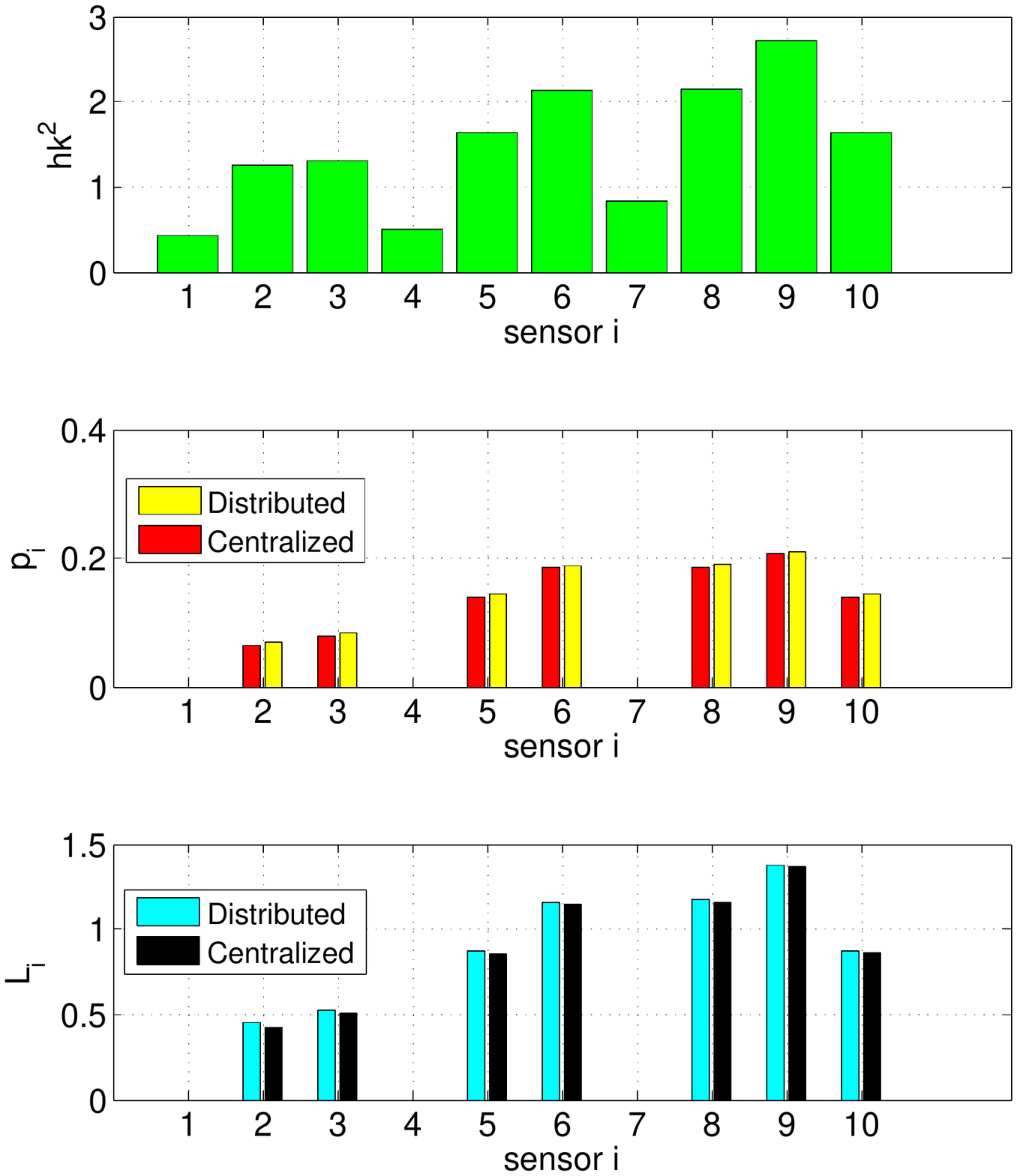}}
    \caption[Centralized and decentralized sensor transmit power and channel bit allocation for $P_{fa}=0.1$, $P_t=1$, $U=3$ and $N=10$]{\label{fig:fig_1}
Centralized and decentralized sensor transmit power and channel bit allocation for $P_{fa}=0.1$, $P_t=1$, $U=3$, $\xi_a=-4$ dB, $N=10$ and $s_i(n)=0.2 \ \forall i$.}
\vspace{0.22cm}
\end{figure}
%
%
%
\begin{figure}[htp!]
 \centerline{\includegraphics[width=80mm ,height=70mm]{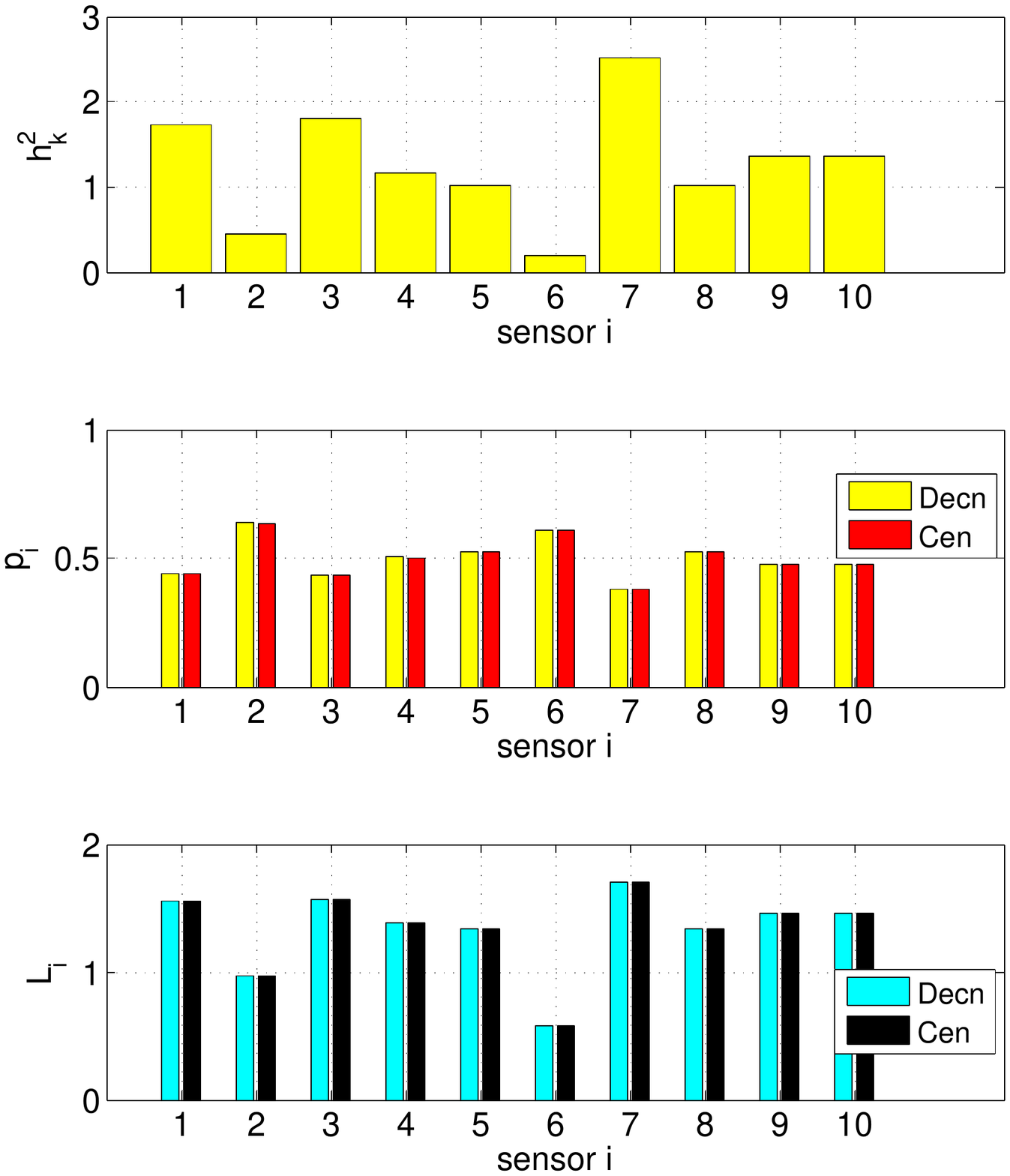}}
    \caption[Disagreement error versus time for $M=10$ sensors.]{\label{fig:fig_4}
    Centralized and decentralized sensor transmit power and channel bit allocation for $P_{fa}=0.1$, $P_t=5$, $U=3$, $\xi_a=-1$ dB, $N=50$ and $s_i(n)=0.3 \ \forall i$.}
\end{figure}
%
%
%
\begin{figure}[htp!]
  \centerline{\includegraphics[width=80mm ,height=70mm]{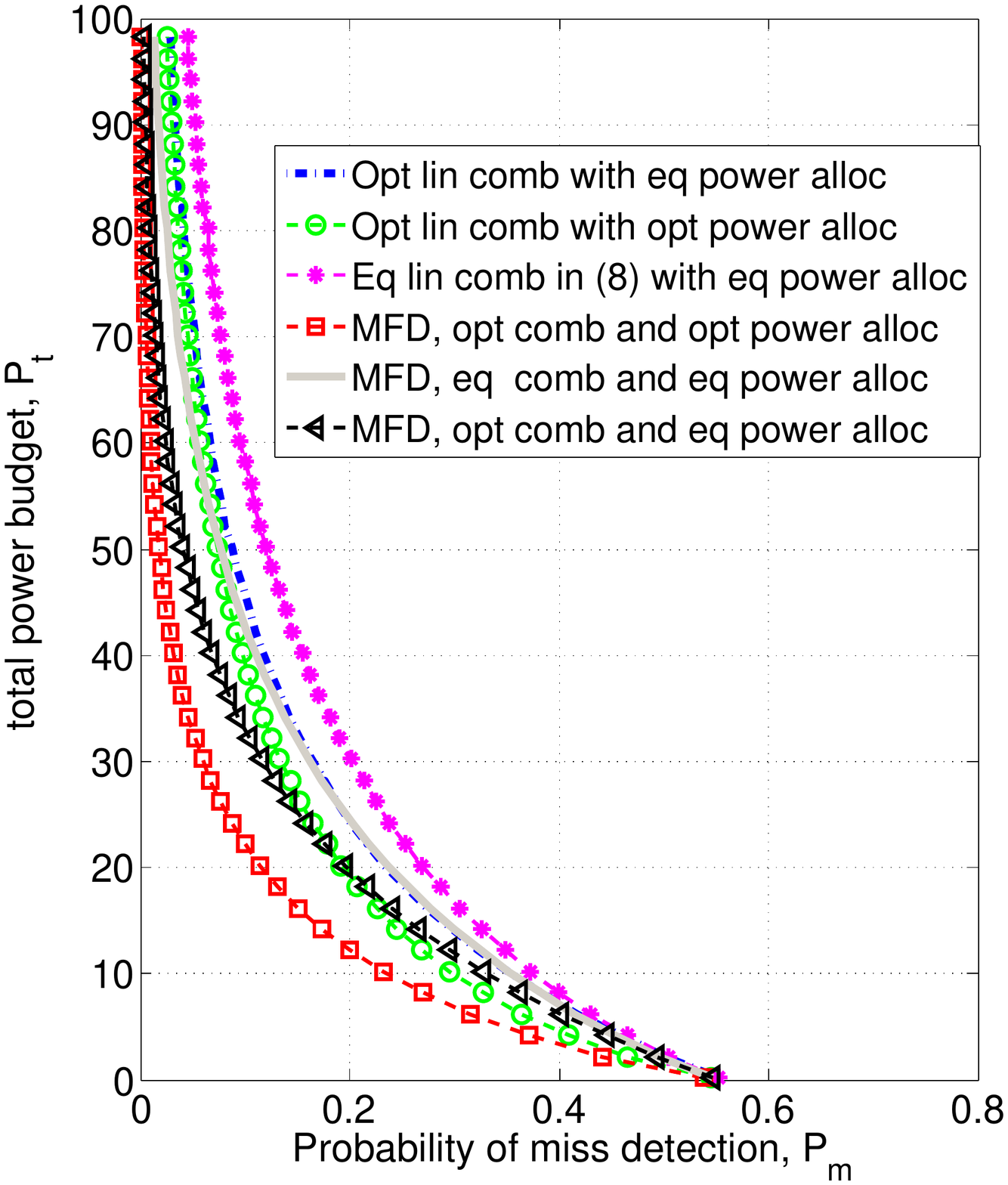}}
    \caption[Disagreement error versus time for $M=10$ sensors.]{\label{fig:fig_9}
    Total power budget ($P_t$) versus probability of mis-detection $(1-P_d)$, with $P_{fa}=0.1$, $U=3$, $\xi_a=-4$ dB, $N=5$ and $M=100$.}
\end{figure}
%
%
%
\begin{figure}[htp!]
  \centerline{\includegraphics[width=80mm ,height=70mm]{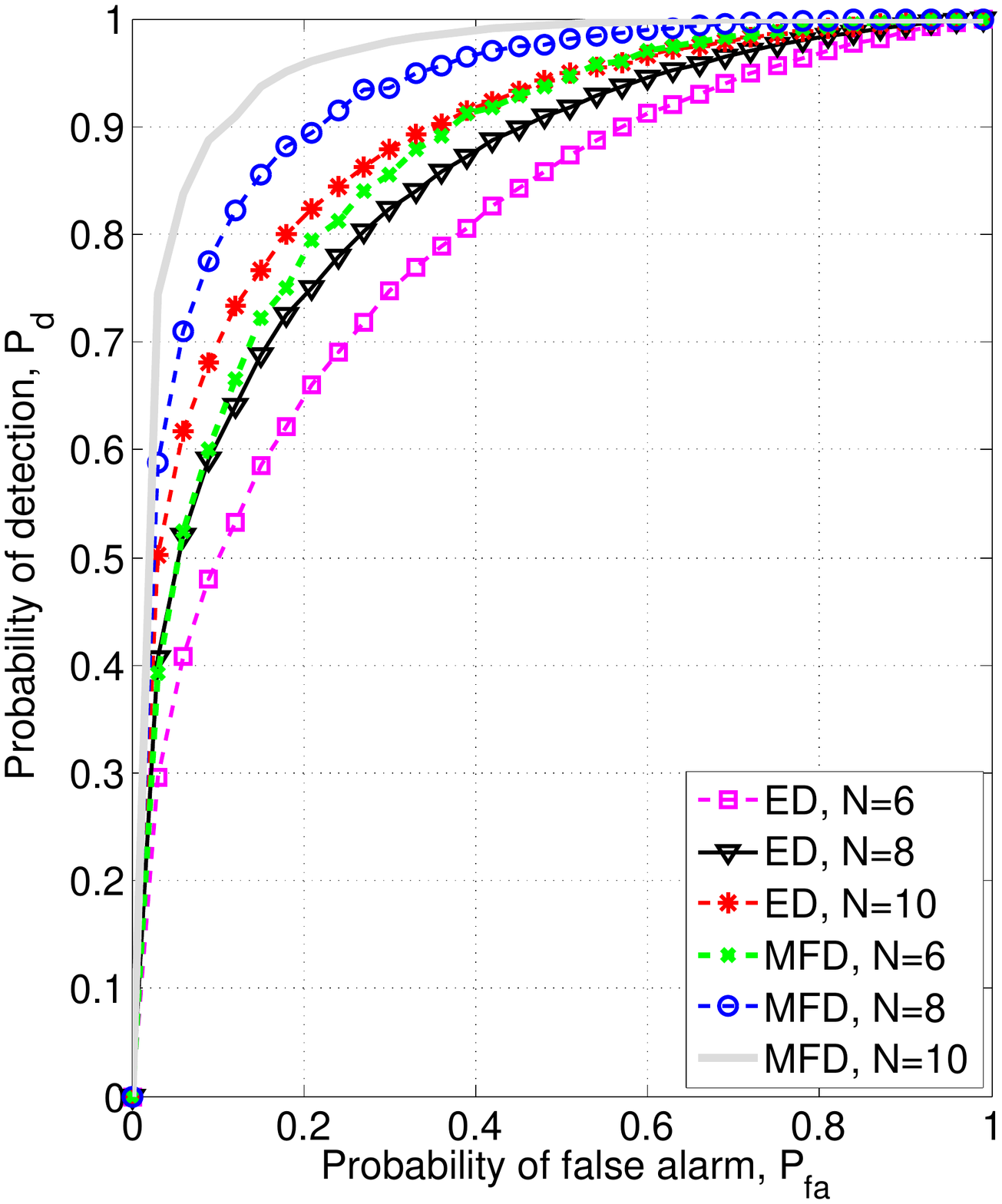}}
    \caption[Disagreement error versus time for $M=10$ sensors.]{\label{fig:fig_10}
Probability of detection ($P_d$) versus probability of false alarm ($P_{fa}$), with $U=3$, $\xi_a=-4$ dB, $P_t=1$ and $M=10$.}
\end{figure}
%
%
%
\begin{figure}[htp!]
  \centerline{\includegraphics[width=80mm ,height=70mm]{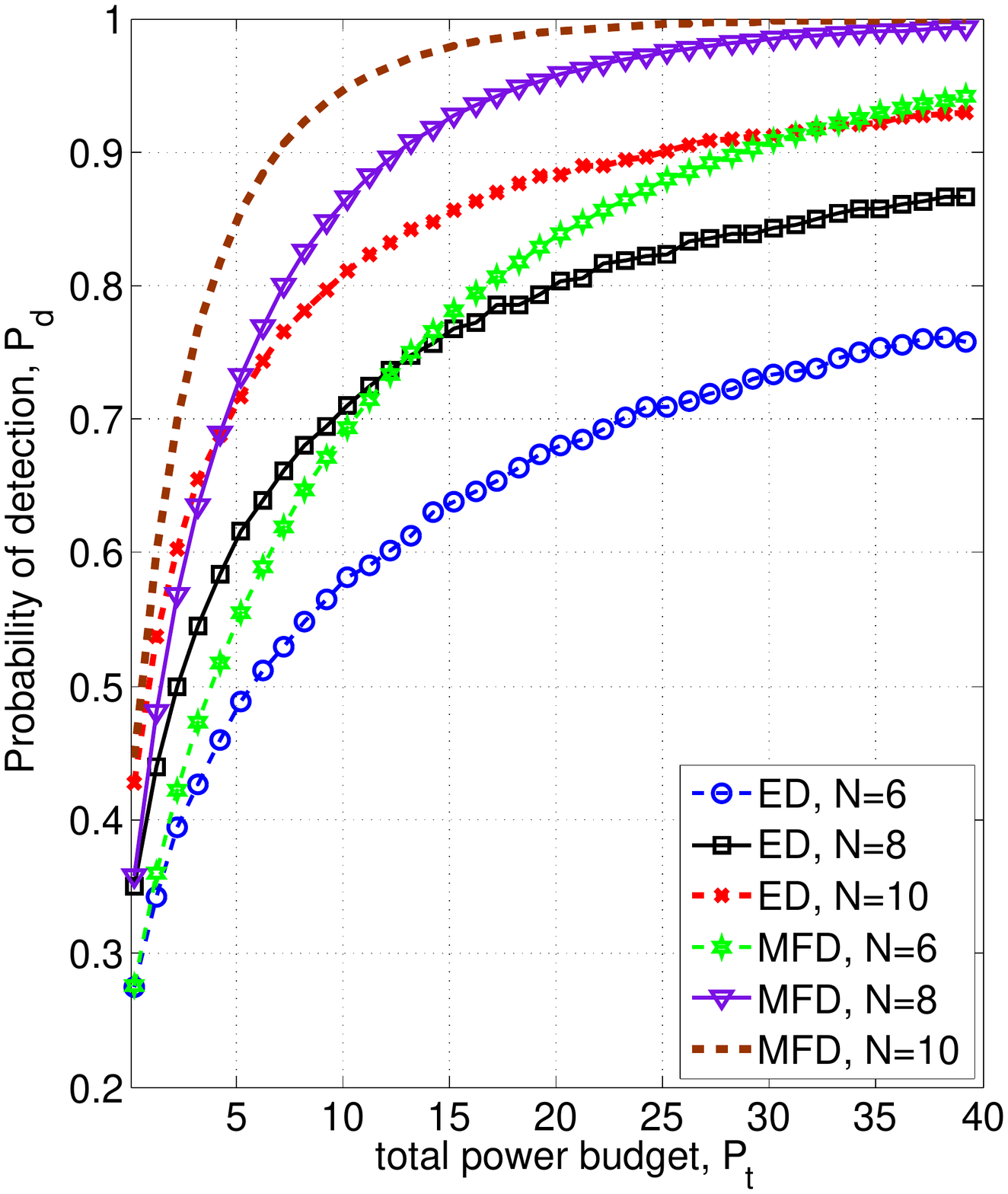}}
   \caption[Disagreement error versus time for $M=10$ sensors.]{\label{fig:fig_11}
 Probability of detection ($P_d$) versus total power budget ($P_{t}$), with $U=3$, $\xi_a=-4$ dB, $P_{fa}=0.1$ and $M=20$.}
\end{figure}
%
%
Finally, we choose $L_i$ with equality in (\ref{eq5}). In Fig. \ref{fig:fig_1}, the middle plot shows  the SN transmit power $p_i$  for the $i^{th}$ SN to the FC channel using two different approaches (i.e., distributed and centralized). The actual channel coefficients (randomly chosen) are in the upper plot in Fig. \ref{fig:fig_1}. Clearly, the performance of our proposed distributed method is very close to the centralized one. As expected, both centralized and decentralized methods  allocate more power to the best channels. In this way, the nodes that have very bad channels (i.e.,nodes that require very high power to transmit) will be censored (i.e., will not transmit even a single bit). 
In Fig. \ref{fig:fig_4}, we show that for large number of samples ($N$) the optimum power allocation scheme tends to a uniform power allocation as expected (see the definition of $\boldsymbol R$ in Section \ref{Decentralized optimum weight combining and power allocation}). Fig. \ref{fig:fig_9} shows the total power budget ($P_t$) against the mis-detection (1-$P_d$) performance for 6 different schemes. The energy detector (ED) performance tends to converge to the matched filter detector for a low power budget ($P_t$). Fig. \ref{fig:fig_10} shows the receiver operating characteristic against the sample number ($N$). As expected, the matched filter detector outperforms the energy detector but it requires full knowledge of the useful signal. And in Fig. \ref{fig:fig_11}, we examine the probability of detection ($P_d$)  performance against the total power budget ($P_t$). As $P_t$ increases, then $P_d$ improves. 
%
%
\section{Conclusion}
\label{CONCLUSION}
We have shown how to perform distributed detection, via SNs transmitting a quantized version of the received energy test statistic to the FC. In addition we have derived the optimal linear combining weights at the FC and proposed a novel distributed algorithm to calculate the optimal transmit power for each SN in order to maximize $P_d$. In this way, the SN can allocate its own transmit power by exchanging information with its own neighbours. What makes this scheme very useful and attractive is that the only value that they should exchange among neighbours is their own transmit power at the current state. The algorithm is robust and easy implementable.
\ifCLASSOPTIONcaptionsoff
  \newpage
\fi

\end{document}